\documentclass[12pt]{article}
\begin{document}
\title{Finite viscoelasticity of filled rubbers: experiments and numerical
simulation}

\author{Aleksey D. Drozdov and Al Dorfmann\\
Institute of Structural Engineering\\
82 Peter Jordan Street\\
1190 Vienna, Austria}
\date{}
\maketitle

\begin{abstract}
Constitutive equations are derived for the viscoelastic behavior
of particle-re\-in\-forc\-ed rubbers at isothermal loading with finite strains.
A filled rubber is thought of as a composite medium where inclusions
with high and low concentrations of junctions between chains
are distributed in a bulk material.
The characteristic size of inhomogeneities is assumed to be small
compared to the dimensions of a specimen and to substantially exceed
the radius of gyration for macromolecules.
The inclusions with high concentration of junctions are associated
with regions of suppressed mobility of chains that surround isolated
clusters of filler and its secondary network.
The regions with low concentration of junctions arise during
the mixing process due to the inhomogeneity in spatial distribution of a cross-linker.
With reference to the theory of transient networks, the viscoelastic
response of elastomers is ascribed to the thermally activated
processes of breakage and reformation of strands in the domains
with low concentration of junctions.
Stress--strain relations for particle-reinforced rubbers
are developed by using the laws of thermodynamics.
Adjustable parameters in the constitutive equations are found by fitting
experimental data in tensile relaxation tests for several grades
of unfilled and carbon black (CB) filled rubber.
It is evidenced that even at moderate finite deformations
(with axial elongations up to 100 \%),
the characteristic rate of relaxation is noticeably affected by strains.
Unlike glassy polymers, where the growth of longitudinal strain results
in an increase in the rate of relaxation, the growth of the elongation
ratio for natural rubber (unfilled or CB reinforced)
implies a decrease in the relaxation rate, which may be explained by partial
crystallization of chains in the regions with low concentration of junctions.
\end{abstract}
\newpage

\section{Introduction}

This paper is concerned with the viscoelastic behavior of unfilled and
particle-reinforced rubbers at isothermal loading with finite strains.
The time-dependent response of rubbery polymers has attracted
substantial attention in the past decade, see, e.g., \cite{GS92}--\cite{HS01},
which may be explained by numerous applications of rubber-like materials
in industry (vehicle tires, shock absorbers, earthquake bearings,
seals, flexible joints, solid propellants, etc.).

Numerous observations in uniaxial and biaxial tests evidence that
rubbers reinforced with carbon black or silica,
as well as solid propellants filled with micrometer-size hard particles
reveal strong time-dependent response which may be associated
with their viscoelastic behavior
\cite{HS95,Lio96,Lio97,BB98,Lio98,SE98,MK00,WL00,HS01}.
A mechanism for stress relaxation at the molecular level remains, however,
an unresolved problem.

Aksel and H\"{u}bner \cite{AH96} carried our tensile relaxation tests
at ambient temperature on polybutadiene rubber filled with glass beads.
Analyzing compounds with various concentrations of filler, they reported
that (i) relaxation was not observed in the unfilled rubber and
(ii) the strength of relaxation (measured as the ratio of the increment
of the longitudinal stress, $\Delta\sigma(t)=\sigma(0)-\sigma(t)$,
at a fixed time $t>0$ to the stress, $\sigma(0)$, at the beginning
of the tests) increased with the content of reinforcement.
As a mechanism for stress relaxation, they mentioned slippage of polymeric chains
along the filler surfaces in the linear regime (at relatively small strains)
and detachment of chains from the surfaces of beads and creation of vacuoles
at large deformations.

Liang et al. \cite{LLT99} presented experimental data in dynamic tensile tests
(frequency 1 Hz) on ternary composite of polypropylene and
EPDM (ethylene propylene diene monomers) elastomer filled with glass beads.
They demonstrated that the mechanical damping (estimated by means of
the loss modulus) of the compound with untreated beads decreased with
the filler content, in contrast with the observations of Ref. \cite{AH96}.

Leblanc and Cartault \cite{LC01} measured dynamic shear response of
uncured styrene--butadiene rubber filled with various amounts
of carbon black and silica and reported that the compounds exhibited
rather limited time-dependent response.
They concluded that cross-links played the key role in the viscoelastic
behavior of rubbers, whereas the effect of filler is relatively weak.

Clarke et al. \cite{CET00} carried out long-term (with the duration over
10$^{6}$ s) tensile relaxation experiments on two unfilled elastomers
(natural polyisoprene rubber with polysulphur cross-links
and synthetic polyacrylate rubber with carbon--carbon
and carbon--oxygen cross-links) at an elevated temperature ($T=80$ $^{\circ}$C).
They reported that at relatively small strains ($\epsilon=0.05$),
natural rubber demonstrated significant mechanical relaxation,
whereas the time-dependent response of the synthetic elastomer
was not revealed.
Slow relaxation in natural rubber was ascribed to isomerisation of
polysulphur cross-links, but no explanation was provided for
the stress relaxation in natural rubber in short-term mechanical tests,
as well as for its absence in the polyacrylate elastomer.

The objective of the present study is to shed some light on the mechanism
of stress relaxation in unfilled and carbon black (CB) filled natural rubbers.
For this purpose, we carry our tensile relaxation tests at room temperature
on specimens with various contents of filler and develop constitutive
equations for the time-dependent response of particle-reinforced elastomers.
Adjustable parameters in the stress--strain relations are found
by fitting observations at various elongation ratios, $\lambda$,
in the region of moderate finite deformations, $\lambda\in (1,2)$.

To describe the time-dependent behavior of rubbers, we employ the concept
of temporary networks.
According to the statistical theory of macromolecules,
a rubbery polymer is thought of as a network of long chains bridged by
junctions (chemical and physical cross-links, entanglements
and van der Waals forces).
A strand between two neighboring junctions is modelled as a set of mers
connected in sequel.
The classical theory of rubber elasticity \cite{Tre75}
treats all junctions as permanent, which implies that strands cannot
be detached from the junctions.
To predict the time-dependent response of rubbery polymers,
Green and Tobolsky \cite{GT46} introduced a concept of temporary junctions
and presumed that chains can slip from the junctions as they are
thermally agitated.
This theory was revised by Yamamoto \cite{Yam56}, Lodge \cite{Lod68}
and Tanaka and Edwards \cite{TE92} and was applied to the description
of the viscoelastic behavior of rubbers
by Drozdov \cite{Dro97} and Ernst and Septanika \cite{ES99}.

According to the Green--Tobolsky concept, a strand whose ends are linked
to separate neighboring junctions is treated as an active one.
Snapping of an end of a strand from a junction is thought of as its
breakage (transition from the active state of the strand to its dangling state).
When a dangling chain captures a junction (which reflects creation of a new
physical cross-link or entanglement), a new active strand
merges with the network.
Breakage and reformation of active strands are treated as thermally
activated processes: attachment and detachment events occur at random times
as they are driven by thermal fluctuations.

A particle-reinforced rubber is modelled as a micro-composite,
where domains with low and high concentrations of junctions
are distributed in a matrix with a ``normal" concentration of
cross-links between chains.
Regions with low concentration of junctions (RLCJ) arise during the
mixing process because of the inhomogeneity in the spatial distribution
of a cross-linker.
Regions with high concentration of junctions (RHCJ) are formed
in a particle-reinforced elastomer in neighborhoods of filler clusters,
where the filler particles and their aggregates play the role of extra
cross-links between polymeric chains.
Breakage and reformation of strands are assumed to take place in the regions
with low concentration of junctions only, whereas in the elastomeric matrix
and in the domains with high concentration of junctions slippage of chains
from cross-links is prevented by surrounding macromolecules.
The viscoelastic response is ascribed to the reformation of strands in RLCJs only,
whereas the host matrix is treated as a nonlinear elastic medium.
We postulate that the response of domains with high concentration of
junctions is substantially affected by mechanical loading, which results in
damage (rupture) of isolated clusters of filler and its secondary network.
Mechanically induced changes in the topology of filler clusters
are described by a vector of internal variables, ${\bf a}$, which
is not concretized in the present study.
To distinguish between the time-dependent response of the composite
in the standard relaxation tests driven by reformation of strands in RLCJs
and that induced by changes in the internal variables, we presume
that the derivative of the vector ${\bf a}$ with respect to time vanishes
when the rate-of-strain tensor equals zero, see Eq. (48).

The aim of this paper is to derive a constitutive model for the time-dependent
behavior of particle-reinforced rubbers and to apply this model to
the analysis of experimental data in tensile relaxation tests.
Our purpose is to assess the effects of filler content and
elongation ratio on the relaxation spectrum and the rate of stress relaxation.

The study is organized as follows.
The free energy density of a particle-reinforced elastomer
is determined in Section 2.
Stress--strain relations at finite strains
are derived in Section 3 by using the laws of thermodynamics.
Uniaxial extension of a specimen is analyzed in Section 4.
Section 5 deals with experimental data in tensile relaxation tests
for three different grades of CB filled rubber.
Adjustable parameters in the constitutive equations
are found in Section 6 by fitting observations.
A brief discussion of our findings is presented in Section 7.
Some concluding remarks are formulated in Section 8.

\section{Free energy of particle-reinforced rubber}

A particle-reinforced rubber is modelled as a micro-composite where inclusions
with high and low concentrations of junctions are randomly distributed
in the bulk material.
The composite is treated as an incompressible and isotropic
elastic medium with the free energy density (per unit mass)
\begin{equation}
\Psi=\Psi^{\circ}+V+W,
\end{equation}
where $\Psi^{\circ}$ is the free energy in the stress-free state
at the reference temperature $T^{\circ}$,
$V$ is the free energy of thermal motion of chains,
and $W$ is the increment of free energy driven by mechanical deformations.

We confine ourselves to quasi-static loadings with moderate finite strains,
when mechanically induced changes in temperature, $\Delta T=T-T^{\circ}$,
are rather weak \cite{KB97}.
This implies that the effect of temperature on the specific heat,
$c$, as well as thermal expansion of the elastomer may be neglected.
Assuming $c$ to be temperature-independent, we adopt the conventional
formula for the free energy of thermal motion
\begin{equation}
V=(c-S^{\circ})(T-T^{\circ})-c T \ln\frac{T}{T^{\circ}},
\end{equation}
where $S^{\circ}$ is the entropy per unit mass in the stress-free state
at the reference temperature.

Let $\mu_{\rm L}$ and $\mu_{\rm H}$ be mass fractions of regions with
low and high concentrations of junctions.
Taking into account that the strain energy density, $W$, is
an extensive thermodynamic potential, we find that
\begin{equation}
W=(1-\mu_{\rm L}-\mu_{\rm H})W_{\rm B}+\mu_{\rm L}W_{\rm L}+\mu_{\rm H}W_{\rm H},
\end{equation}
where $W_{\rm B}$ is the mechanical energy per unit mass of the bulk medium,
and $W_{\rm L}$, $W_{\rm H}$ are the strain energy densities per unit mass
for regions with low and high concentrations of junctions, respectively.

\subsection{Mechanical energy of the host matrix}

The bulk medium is treated as a permanent network of long chains
bridged by junctions (chemical cross-links, physical cross-links
whose life-time exceeds the duration of a mechanical test, and
entanglements). We adopt the affinity hypothesis which neglects
thermal oscillations of junctions and postulates that
the displacements of junctions at the micro-scale coincide with
the displacements of appropriate points of a sample at the macro-level.

To clarify this assumption, we consider a strand between two
contiguous junctions $A_{1}$ and $A_{2}$.
In the stress-free state, the junctions occupy points with the radius vectors ${\bf
R}_{1}^{\circ}$ and ${\bf R}_{2}^{\circ}$.
Denote by ${\bf R}_{1}(t)$ and ${\bf R}_{2}(t)$ radius vectors of these junctions
at an arbitrary instant $t\geq 0$ (the time $t=0$ corresponds to
the instant when external loads are applied to the specimen).
Let ${\bf r}^{\circ}={\bf R}_{2}^{\circ}-{\bf R}_{1}^{\circ}$ be the
end-to-end vector for the strand in the stress-free state and
${\bf r}(t)={\bf R}_{2}(t)-{\bf R}_{1}(t)$ the end-to-end vector
in the deformed state at time $t$.
According to the affinity hypothesis, the deformation gradient
${\bf \nabla}_{{\bf r}^{\circ}} {\bf r}(t)$
for motion of junctions coincides with the deformation gradient
${\bf \nabla}_{{\bf R}^{\circ}} {\bf R}(t)={\bf \nabla}^{\circ} {\bf R}(t)$
for macro-deformation.
This implies that the Cauchy deformation tensor ${\bf c}(t)$ for the
network equals the Cauchy deformation tensor for macro-deformations,
\[
{\bf C}^{\circ}(t)={\bf \nabla}^{\circ} {\bf R}(t)\cdot
\Bigl [{\bf \nabla}^{\circ} {\bf R}(t)\Bigr ]^{\top},
\]
and the rate-of-strain tensor for the network ${\bf d}(t)$
coincides with the rate-of-strain tensor for macro-deformations
\begin{equation}
{\bf D}(t)=\frac{1}{2}\Bigl [ {\bf \nabla}(t) {\bf V}(t)+ \Bigl
({\bf \nabla}(t){\bf V}(t)\Bigr )^{\top}\Bigr ].
\end{equation}
Here ${\bf \nabla}(t)={\bf \nabla}_{{\bf R}(t)}$ is the gradient
operator in the deformed state of the specimen,
\[
{\bf V}(t)=\frac{d {\bf R}}{dt}(t)
\]
is the velocity vector,
the dot stands for inner product,
and $\top$ denotes transpose.

For an isotropic incompressible medium, the strain energy density $W_{\rm B}$
depends on the current temperature $T$
(the strong effect of temperature reflects the entropic nature
of the mechanical energy for elastomers) and the first two principal invariants
of the Cauchy deformation tensor ${\bf C}^{\circ}$,
\begin{equation}
W_{\rm B}(t)=\bar{W}_{\rm B}\Bigl (T(t), I_{1}({\bf C}^{\circ}(t)),
I_{2}({\bf C}^{\circ}(t))\Bigr ).
\end{equation}
Disregarding thermally induced changes in the reference state
(at small increments of temperature, $\Delta T$, of the order of several K \cite{KB97}
thermal expansion of rubber is negligible compared to the strains
under consideration), we assume the function $\bar{W}_{\rm B}$
to obey the condition
\begin{equation}
\bar{W}_{\rm B}(T,I_{1},I_{2})\Bigl |_{I_{1}=3, I_{2}=3}=0,
\end{equation}
which means that the strain energy vanishes in the stress-free state.

The mechanical energy per strand of the bulk material is given by
\begin{equation}
w_{\rm B}(t)=\frac{W_{\rm B}(t)}{N_{\rm B}},
\end{equation}
where $N_{\rm B}$ is the average number of strands per unit mass.

\subsection{Mechanical energy of domains with low concentration of junctions}

Regions with low concentration of junctions are assumed to be randomly distributed
in the host matrix as inclusions with various sizes and shapes.
Any RLCJ is modelled as a transient network of long chains,
where breakage and reformation events occur at random times
as active and dangling strands are agitated by thermal fluctuations.
With reference to the theory of thermally activated processes \cite{Eyr36},
we assume that the rate of breakage for a strand in a stress-free
medium is given by
\begin{equation}
\Gamma=\Gamma_{0} \exp \Bigl (-\frac{\tilde{\omega}}{k_{\rm B}T} \Bigr ),
\end{equation}
where $\tilde{\omega}$ is the energy for breakage,
$k_{\rm B}$ is Boltzmann's constant,
and the pre-factor $\Gamma_{0}$ weakly depends on temperature.
It follows from Eq. (8) that for any $\tilde{\omega}>0$,
$\Gamma$ vanishes at low temperatures, $T\to 0$,
and $\Gamma$ tends to $\Gamma_{0}$ at high temperatures, $T\to\infty$.
This implies that the coefficient $\Gamma_{0}$ may be associated with
the breakage rate at elevated temperatures.

We introduce the dimensionless energy for breakage,
\[ \omega=\frac{\tilde{\omega}}{k_{\rm B}T^{\circ}}, \]
and neglect the effect of small increments of temperature,
$\Delta T$, on the rate of breakage.
This results in the formula for the rate of transition from the active
state of a strand to its dangling state
\begin{equation}
\Gamma(\omega)=\Gamma_{0} \exp (-\omega).
\end{equation}
Equation (9) is assumed to be satisfied for an arbitrary loading,
provided that the coefficient $\Gamma$ depends on the Cauchy deformation
measure, ${\bf C}^{\circ}$, and additional internal variables that characterize
damage of the secondary network and isolated clusters of filler.
Because a detailed description of damage for particle-reinforced composites
is beyond the scope of the present study, we accept a formal standpoint and
postulate that rupture of filler clusters is described by a vector of
internal parameters ${\bf a}$.
This implies that for an isotropic and incompressible medium,
\begin{equation}
\Gamma_{0}=\Gamma_{0}\Bigl ({\bf a},I_{1}({\bf C}^{\circ}),I_{2}({\bf C}^{\circ})\Bigr ),
\end{equation}
where the function $\Gamma_{0}({\bf a},I_{1},I_{2})$ is to be found by
fitting experimental data.
It follows from Eqs. (9) and (10) that
\begin{equation}
\Gamma(t,\omega)=\Gamma_{\ast}(t) \exp (-\omega)
\end{equation}
with
\[
\Gamma_{\ast}(t)=\Gamma_{0}\Bigl ({\bf a}(t),I_{1}({\bf C}^{\circ}(t)),I_{2}({\bf C}^{\circ}(t))\Bigr ).
\]
Denote by $p(\omega)$ the distribution function for energies of breakage
for strands in RLCJs.
The function $p(\omega)$ is assumed to be robust with respect
to the influence of thermo-mechanical factors.
This means that $p$ is treated as a function which is temperature-independent
(for small increments of temperature, $\Delta T$) and which does not
depend on the history of loading.
To fit experimental data, we use the quasi-Gaussian distribution
\begin{equation}
p(\omega) = p_{0}\exp \biggl [-\frac{(\omega-\Omega)^{2}}{2\Sigma^{2}}\biggr ],
\end{equation}
where $\Omega$ and $\Sigma$ are adjustable parameters, and the constant $p_{0}$
is found from the condition
\begin{equation}
\int_{0}^{\infty} p(\omega) d\omega =1.
\end{equation}
Let $dN(\omega)$ be the concentration of strands in RLCJs (per unit mass)
whose energy for breakage, $\omega^{\prime}$, is located
in the interval $[\omega,\omega +d\omega]$.
The quantity $dN(\omega)$ reads
\[ dN(\omega)=N_{\rm L} p(\omega)d\omega , \]
where $N_{\rm L}$ is the average number of strands per unit mass of RLCJs.

The kinetics of reformation is described by the function $X(t,\tau,\omega)$,
which equals the current (at time $t$) number (per unit mass) of active
(merged to the network) strands with the energy for breakage $\omega$
that have last been bridged to the network before an instant $\tau\in [0,t]$.
The function $X$ entirely determines the current state of active strands
in the regions with low concentration of junctions.
The quantity $X(t,t,\omega)$ equals the number of active strands
(per unit mass) with the energy for breakage $\omega$ at time $t$.
In particular, $X(0,0,\omega)$ is the initial number of active strands
with the energy for breakage $\omega$,
\begin{equation}
X(0,0,\omega)=N_{\rm L} p(\omega).
\end{equation}
The amount
\[ \frac{\partial X}{\partial \tau}(t,\tau,\omega) \biggl |_{t=\tau} \; d\tau \]
is the number (per unit mass) of active strands with the energy for breakage $\omega$
that have been linked to the network within the interval $[\tau,\tau+d\tau ]$.
The quantity
\[ \frac{\partial X}{\partial \tau} (t,\tau,\omega)\;d\tau \]
determines the number of these strands that have not been broken
during the interval $[\tau, t]$.
The amount
\[ -\frac{\partial X}{\partial t} (t,0,\omega)\;dt \]
is the number of active strands (per unit mass) that detach from the
network (for the first time) within the interval $[t,t+dt ]$,
and the quantity
\[ -\frac{\partial^{2} X}{\partial t\partial \tau} (t,\tau,\omega)\;dtd\tau \]
is the number of strands (per unit mass) that have last been linked
to the network within the interval $[\tau,\tau+d\tau ]$
and leave the network (for the first time after merging)
during the interval $[t,t+dt ]$.

The kinetics of evolution for a transient network is determined
by the relative rate of breakage for active strands, $\Gamma(t,\omega)$,
and by the rate of merging with the network for dangling strands, $\gamma(t,\omega)$.
The assumption that $\Gamma$ and $\gamma$ are functions of energy
for breakage, $\omega$, distinguishes the present model from previous ones.
The dependence of the rates of breakage and reformation on time, $t$,
reflects the effects of current strains and the damage variables, ${\bf a}$,
on these parameters under time-dependent loading.

The relative rate of breakage, $\Gamma$, equals the ratio of the number
of active strands broken per unit time to the total number of active strands.
Applying this definition to the strands arisen before loading
and remaining active at the current time $t$, we find that
\begin{equation}
\Gamma(t,\omega) = -\frac{1}{X(t,0,\omega)}\frac{\partial X}{\partial t}(t,0,\omega).
\end{equation}
Applying this definition to the strands merging with the network at time $\tau$
and remaining active at time $t\geq \tau$, we obtain
\begin{equation}
\Gamma(t,\omega) = -\biggl [ \frac{\partial X}{\partial \tau}(t,\tau,\omega) \biggr ]^{-1}
\frac{\partial^{2} X}{\partial t\partial \tau}(t,\tau,\omega).
\end{equation}
In agreement with conventional theories of temporary networks \cite{TE92},
the rate of reformation, $\gamma$, is defined as the number
of dangling strands (per unit mass) bridged to the network per unit time
\begin{equation}
\gamma(t,\omega)=\frac{\partial X}{\partial \tau}(t,\tau,\omega)\biggl |_{\tau=t}.
\end{equation}
We confine ourselves to a model with two possible states (active and dangling) for a strand.
A dangling chain is treated as a chain where stresses totally relax
after its detachment (no memory about previous deformations),
whereas an active one is a chain that preserves the entire memory
about its state at the instant when it merges with the network.
One can hypothesize about an intermediate state between these two
extremities, when the interval between the detachment event and
the subsequent attachment to the network is too small for total relaxation
of stresses, but is sufficiently large to ensure that a part of these stresses
do relax \cite{Pal97}.
This is equivalent to the assumption that a partial memory preserves
in a strand about the history of its deformation.
To simplify the analysis, this case is excluded from our consideration.

Formulas (15) and (16) can be treated as ordinary differential equations for the function $X$.
Integration of these equations with the initial conditions (14) and (17) implies that
\begin{eqnarray}
X(t,0,\omega) &=& N_{\rm L} p(\omega)\exp \biggl [ -\int_{0}^{t} \Gamma(s,\omega) ds\biggr ],
\nonumber\\
\frac{\partial X}{\partial \tau}(t,\tau,\omega) &=&
\gamma(\tau,\omega)\exp \biggl [-\int_{\tau}^{t} \Gamma(s,\omega) ds \biggr ].
\end{eqnarray}
To exclude the rate of reformation, $\gamma$, from Eq. (18),
we postulate that the concentrations of active and dangling
strands are independent of macro-strains.
This means that the number of strands broken per unit time
coincides with the number of strands merging with the network
within the same interval.
The number of strands detached from the network during the interval
$[t,t+dt ]$ equals the sum of the number of initial strands
(not broken until time $t$) that slip from temporary junctions
\[ -\frac{\partial X}{\partial t}(t,0,\omega)\; dt \]
and the number of strands linked with the network within the
interval $[\tau, \tau+d\tau ]$ (for the last time before instant $t$)
for various instants $\tau$ and broken within the interval $[t,t+dt ]$
\[ -\frac{\partial^{2} X}{\partial t\partial \tau}(t,\tau,\omega)\;dt d\tau. \]
This number coincides with the number of strands, $\gamma(t,\omega)\; dt$,
merged with the network within the interval $[t,t+dt]$,
which results in the balance law
\[ \gamma(t,\omega)=-\frac{\partial X}{\partial t}(t,0,\omega)
-\int_{0}^{t} \frac{\partial^{2} X}{\partial t\partial \tau}
(t,\tau,\omega) d\tau . \]
Substitution of Eq. (18) into this equality implies that
\[ \gamma(t,\omega) = \Gamma (t,\omega)\biggl \{
N_{\rm L} p(\omega)\exp \biggl [-\int_{0}^{t} \Gamma(s,\omega) ds \biggr ]
+\int_{0}^{t} \gamma(\tau,\omega)
\exp \biggl [-\int_{\tau}^{t} \Gamma(s,\omega) ds \biggr ] d\tau\biggr \}. \]
To solve this equation, we introduce the notation
\begin{equation}
\tilde{\gamma}(t,\omega)=\gamma(t,\omega)\exp \biggl [\int_{0}^{t}
\Gamma(s,\omega) ds\biggr ],
\end{equation}
and find that
\begin{equation}
\tilde{\gamma}(t,\omega)=\Gamma(t,\omega)\biggl [ N_{\rm L} p(\omega)
+\int_{0}^{t}\tilde{\gamma}(\tau,\omega)d\tau \biggr ].
\end{equation}
Setting $t=0$ in Eq. (20), we obtain
\begin{equation}
\tilde{\gamma}(0,\omega)=N_{\rm L} \Gamma(0,\omega)p(\omega).
\end{equation}
Differentiation of Eq. (20) with respect to time results in
\[ \frac{\partial \tilde{\gamma}}{\partial t}(t,\omega)
=\biggl [ \Gamma(t,\omega)+\frac{1}{\Gamma(t,\omega)}
\frac{\partial \Gamma}{\partial t}(t,\omega)\biggr ]
\tilde{\gamma}(t,\omega). \]
Integration of this equality from zero to $t$ yields
\begin{equation}
\ln \frac{\tilde{\gamma}(t,\omega)}{\tilde{\gamma}(0,\omega)}
=\ln \frac{\Gamma(t,\omega)}{\Gamma(0,\omega)}
+\int_{0}^{t}\Gamma(s,\omega) ds.
\end{equation}
It follows from Eqs. (21) and (22) that
\[ \tilde{\gamma}(t,\omega)=N_{\rm L} p(\omega)\Gamma(t,\omega)
\exp \biggl [\int_{0}^{t} \Gamma(s,\omega) ds \biggr ]. \]
Combining this equality with Eq. (19), we find the rate of reformation
\begin{equation}
\gamma(t,\omega)=N_{\rm L} p(\omega)\Gamma(t,\omega).
\end{equation}
Denote be  $w_{\rm L}$ the mechanical energy of an active strand
in domains with low concentration of junctions.
Because RLCJs are treated as isotropic and incompressible media,
$w_{\rm L}$ is assumed to depend on temperature, $T$,
and the first two principal invariants of the Cauchy deformation
tensor.
The conventional hypothesis that stress in a dangling strand
totally relaxes before this strand captures a new junction implies
that the reference (stress-free) state of a strand merging with the network
at time $\tau$ coincides with the deformed state of the network at that instant.
This means that for a strand not broken within the interval $[0,t]$,
\begin{equation}
w_{\rm L}(t,0)=\bar{w}_{\rm L}\Bigl (T(t), I_{1}({\bf C}^{\circ}(t)),
I_{2}({\bf C}^{\circ}(t))\Bigr ),
\end{equation}
and for an active strand which has last been reformed at time $\tau\in [0,t]$,
\begin{equation}
w_{\rm L}(t,\tau)=\bar{w}_{\rm L}\Bigl (T(t), I_{1}({\bf C}(t,\tau)),I_{2}({\bf C}(t,\tau))\Bigr ),
\end{equation}
where ${\bf C}(t,\tau)$ is the relative Cauchy deformation tensor for transition
from the deformed state of a specimen at instant $\tau$ (when the strand
has last been merged to the network) to its deformed state at time $t\geq \tau$.
The tensor ${\bf C}(t,\tau)$ is expressed in terms of the deformation gradient
${\bf \nabla}(\tau){\bf R}(t)$ by the formula
\[
{\bf C}(t,\tau)={\bf \nabla}(\tau) {\bf R}(t)\cdot
\Bigl [{\bf \nabla}(\tau) {\bf R}(t)\Bigr ]^{\top}.
\]
Summing mechanical energies for strands with various energies for breakage, $\omega$,
merged with the network at various instants, $\tau\in [0,t]$,
and neglecting the energy of interaction between strands,
we arrive at the strain energy density per unit mass of RLCJs
\begin{equation}
W_{\rm L}(t)=\int_{0}^{\infty} \biggl [ X(t,0,\omega)w_{\rm L}(t,0)
+\int_{0}^{t} \frac{\partial X}{\partial \tau}(t,\tau,\omega)w_{\rm L}(t,\tau)
d\tau \biggr ] d\omega.
\end{equation}
Substitution of expressions (18), (23), (24) and (25) into Eq. (26) implies that
\begin{eqnarray}
W_{\rm L}(t) &=& \int_{0}^{\infty} p(\omega)d\omega
\biggl \{ \bar{W}_{\rm L}\Bigl (T(t), I_{1}({\bf C}^{\circ}(t)),
I_{2}({\bf C}^{\circ}(t))\Bigr )\exp \Bigl [-\int_{0}^{t} \Gamma(s,\omega) ds\Bigr ]
\nonumber\\
&&+\int_{0}^{t} \Gamma(\tau,\omega)
\bar{W}_{\rm L}\Bigl (T(t), I_{1}({\bf C}(t,\tau)),I_{2}({\bf C}(t,\tau))\Bigr )
\nonumber\\
&&\times
\exp \Bigl [-\int_{\tau}^{t} \Gamma(s,\omega) ds\Bigr ] d\tau \biggr \},
\end{eqnarray}
where, in accord with Eq. (7), we use the notation
\[ \bar{W}_{\rm L}(T, I_{1},I_{2})=N_{\rm L}\bar{w}_{\rm L}(T, I_{1},I_{2}). \]

\subsection{Mechanical energy of domains with high concentration of junctions}

Because the present study focuses on the viscoelastic response of
filled rubbers, we do not dwell on changes in the mechanical energy
of regions with high concentration of junctions driven by rupture
of the secondary network and isolated clusters of filler.
It is assumed that the strain energy density per unit mass of RHCJs reads
\begin{equation}
W_{\rm H}(t)=\bar{W}_{\rm H}\Bigl ( {\bf a}(t),T(t), I_{1}({\bf C}^{\circ}(t)),
I_{2}({\bf C}^{\circ}(t))\Bigr ),
\end{equation}
where no concrete form is provided for the function
$\bar{W}_{\rm H}({\bf a},T,I_{1},I_{2})$.

\subsection{Mechanical energy of the compound}

Substitution of expressions (5), (27) and (28) into Eq. (3)
results in the formula
\begin{eqnarray}
W(t)&=& (1-\mu_{\rm L}-\mu_{\rm H})\bar{W}_{\rm B}\Bigl (T(t), I_{1}({\bf C}^{\circ}(t)),
I_{2}({\bf C}^{\circ}(t))\Bigr )
\nonumber\\
&&+\mu_{\rm L} \int_{0}^{\infty} p(\omega)d\omega
\biggl \{ \bar{W}_{\rm L}\Bigl (T(t), I_{1}({\bf C}^{\circ}(t)),
I_{2}({\bf C}^{\circ}(t))\Bigr )\exp \Bigl [-\int_{0}^{t} \Gamma(s,\omega) ds\Bigr ]
\nonumber\\
&&+\int_{0}^{t} \Gamma(\tau,\omega)
\bar{W}_{\rm L}\Bigl (T(t), I_{1}({\bf C}(t,\tau)),I_{2}({\bf C}(t,\tau))\Bigr )
\exp \Bigl [-\int_{\tau}^{t} \Gamma(s,\omega) ds\Bigr ] d\tau \biggr \}
\nonumber\\
&&+\mu_{\rm H}\bar{W}_{\rm H}\Bigl ( {\bf a}(t),T(t), I_{1}({\bf C}^{\circ}(t)),
I_{2}({\bf C}^{\circ}(t))\Bigr ).
\end{eqnarray}
Bearing in mind Eq. (6), we find from Eq. (29) that
\begin{equation}
\frac{dW}{dt}(t)=J(t)\frac{dT}{dt}(t)+{\bf A}(t)\cdot\frac{d{\bf a}}{dt}(t)
+U_{1}^{\circ}(t)\frac{dI_{1}}{dt}({\bf C}^{\circ}(t))
+U_{2}^{\circ}(t)\frac{dI_{2}}{dt}({\bf C}^{\circ}(t))
+U(t)-Y(t).
\end{equation}
Here we introduce the notation
\begin{eqnarray}
J(t)&=& (1-\mu_{\rm L}-\mu_{\rm H})\frac{\partial \bar{W}_{\rm B}}{\partial T}
\Bigl (T(t), I_{1}({\bf C}^{\circ}(t)), I_{2}({\bf C}^{\circ}(t))\Bigr )
\nonumber\\
&&+\mu_{\rm L} \int_{0}^{\infty} p(\omega)d\omega
\biggl \{ \frac{\partial \bar{W}_{\rm L}}{\partial T}
\Bigl (T(t), I_{1}({\bf C}^{\circ}(t)),I_{2}({\bf C}^{\circ}(t))\Bigr )
\exp \Bigl [-\int_{0}^{t} \Gamma(s,\omega) ds\Bigr ]
\nonumber\\
&&+\int_{0}^{t} \Gamma(\tau,\omega)\frac{\partial \bar{W}_{\rm L}}{\partial T}
\Bigl (T(t), I_{1}({\bf C}(t,\tau)),I_{2}({\bf C}(t,\tau))\Bigr )
\exp \Bigl [-\int_{\tau}^{t} \Gamma(s,\omega) ds\Bigr ] d\tau \biggr \}
\nonumber\\
&&+\mu_{\rm H}\frac{\partial \bar{W}_{\rm H}}{\partial T}
\Bigl ( {\bf a}(t),T(t), I_{1}({\bf C}^{\circ}(t)),I_{2}({\bf C}^{\circ}(t))\Bigr ),
\nonumber\\
{\bf A}(t) &=& \mu_{\rm H}{\bf \nabla}_{\bf a}
\bar{W}_{\rm H}\Bigl ( {\bf a}(t),T(t), I_{1}({\bf C}^{\circ}(t)),I_{2}({\bf C}^{\circ}(t))\Bigr ),
\nonumber\\
U_{k}^{\circ}(t)&=& (1-\mu_{\rm L}-\mu_{\rm H})\frac{\partial \bar{W}_{\rm B}}{\partial I_{k}}
\Bigl (T(t), I_{1}({\bf C}^{\circ}(t)), I_{2}({\bf C}^{\circ}(t))\Bigr )
\nonumber\\
&&+\mu_{\rm L} \int_{0}^{\infty} \frac{\partial \bar{W}_{\rm L}}{\partial I_{k}}
\Bigl (T(t), I_{1}({\bf C}^{\circ}(t)),I_{2}({\bf C}^{\circ}(t))\Bigr )
\exp \Bigl [-\int_{0}^{t} \Gamma(s,\omega) ds\Bigr ]p(\omega)d\omega
\nonumber\\
&&+\mu_{\rm H}\frac{\partial \bar{W}_{\rm H}}{\partial I_{k}}
\Bigl ( {\bf a}(t),T(t), I_{1}({\bf C}^{\circ}(t)),I_{2}({\bf C}^{\circ}(t))\Bigr ),
\nonumber\\
U(t) &=& \mu_{\rm L} \int_{0}^{\infty} p(\omega)d\omega
\int_{0}^{t} \Gamma(\tau,\omega) \biggl [
\frac{\partial \bar{W}_{\rm L}}{\partial I_{1}}
\Bigl (T(t), I_{1}({\bf C}(t,\tau)),I_{2}({\bf C}(t,\tau))\Bigr )
\frac{\partial I_{1}}{\partial t}({\bf C}(t,\tau))
\nonumber\\
&& +\frac{\partial \bar{W}_{\rm L}}{\partial I_{2}}
\Bigl (T(t), I_{1}({\bf C}(t,\tau)),I_{2}({\bf C}(t,\tau))\Bigr )
\frac{\partial I_{2}}{\partial t}({\bf C}(t,\tau))\biggr ]
\exp \Bigl [-\int_{\tau}^{t} \Gamma(s,\omega) ds\Bigr ] d\tau,
\nonumber\\
Y(t) &=& \mu_{\rm L} \int_{0}^{\infty} \Gamma(t,\omega) p(\omega)
\biggl \{ \bar{W}_{\rm L}\Bigl (T(t), I_{1}({\bf C}^{\circ}(t)),
I_{2}({\bf C}^{\circ}(t))\Bigr )\exp \Bigl [-\int_{0}^{t} \Gamma(s,\omega) ds\Bigr ]
\nonumber\\
&&+ \int_{0}^{t} \Gamma(\tau,\omega)
\bar{W}_{\rm L}\Bigl (T(t), I_{1}({\bf C}(t,\tau)),I_{2}({\bf C}(t,\tau))\Bigr )
\nonumber\\
&&\times
\exp \Bigl [-\int_{\tau}^{t} \Gamma(s,\omega) ds\Bigr ] d\tau \biggr \}d\omega.
\end{eqnarray}
The principal invariants of the Cauchy deformation tensor, ${\bf C}(t,\tau)$,
coincide with the principal invariants of the Finger tensor for transition
from the deformed state at time $\tau$ to the deformed state at time $t$,
\[ {\bf F}(t,\tau)=\Bigl [{\bf \nabla}(\tau){\bf R}(t)\Bigr ]^{\top}
\cdot {\bf \nabla}(\tau){\bf R}(t). \]
This assertion implies that
\begin{equation}
\frac{\partial I_{k}}{\partial t}({\bf C}(t,\tau))
=\frac{\partial I_{k}}{\partial t}({\bf F}(t,\tau))
=\frac{\partial I_{k}}{\partial {\bf F}}({\bf F}(t,\tau)):
\biggl [ \frac{\partial {\bf F}}{\partial t}(t,\tau)\biggr ]^{\top},
\end{equation}
where the colon stands for convolution of tensors.
The derivatives of the principal invariants of the Finger tensor read \cite{Dro96}
\[ \frac{\partial I_{1}}{\partial {\bf F}}({\bf F})={\bf I},
\qquad
\frac{\partial I_{2}}{\partial {\bf F}}({\bf F})
=I_{1}({\bf F}){\bf I}-{\bf F}^{\top}, \]
where ${\bf I}$ is the unit tensor.
The derivative of the tensor ${\bf F}$ with respect to time is given by \cite{Dro96}
\begin{equation}
\frac{\partial {\bf F}}{\partial t}(t,\tau)
=\Bigl [ {\bf \nabla}(t){\bf V}(t)\Bigr ]^{\top}
\cdot {\bf F}(t,\tau) +{\bf F}(t,\tau)\cdot{\bf \nabla}(t){\bf V}(t).
\end{equation}
It follows from Eqs. (32) and (33) that
\begin{eqnarray}
\frac{\partial I_{1}}{\partial t}({\bf C}(t,\tau))
&=& 2{\bf F}(t,\tau):{\bf D}(t),
\nonumber\\
\frac{\partial I_{2}}{\partial t}({\bf C}(t,\tau))
&=& 2\Bigl [ I_{1}({\bf C}(t,\tau)){\bf F}(t,\tau)
-{\bf F}^{2}(t,\tau)\Bigr ]:{\bf D}(t),
\end{eqnarray}
where the rate-of-strain tensor ${\bf D}$ is determined by Eq. (4).
By analogy with Eq. (34), we find that
\begin{eqnarray}
\frac{d I_{1}}{d t}({\bf C}^{\circ}(t))
&=& 2{\bf F}^{\circ}(t):{\bf D}(t),
\nonumber\\
\frac{d I_{2}}{d t}({\bf C}^{\circ}(t))
&=& 2\Bigl [ I_{1}({\bf C}^{\circ}(t)){\bf F}^{\circ}(t)
-\Bigl( {\bf F}^{\circ}(t)\Bigr )^{2}\Bigr ]:{\bf D}(t),
\end{eqnarray}
where
\[ {\bf F}^{\circ}(t)=\Bigl [{\bf \nabla}^{\circ} {\bf R}(t)\Bigr ]^{\top}
\cdot {\bf \nabla}^{\circ} {\bf R}(t). \]
Substitution of expression (34) into formula (31) for the functional $U(t)$
implies that
\begin{equation}
U(t)= 2\mu_{\rm L}{\bf \Lambda}_{\rm L}(t):{\bf D}(t).
\end{equation}
Here
\begin{eqnarray}
{\bf \Lambda}_{\rm L}(t) &=& \int_{0}^{\infty} p(\omega)d\omega
\int_{0}^{t} \Bigl [ \phi_{\rm L\,1}(t,\tau){\bf F}(t,\tau)
+\phi_{\rm L\,2}(t,\tau){\bf F}^{2}(t,\tau)\Bigr ]
\nonumber\\
&&\times \Gamma(\tau,\omega)
\exp \Bigl [-\int_{\tau}^{t} \Gamma(s,\omega) ds\Bigr ] d\tau,
\end{eqnarray}
and the functions $\phi_{{\rm L}\,k}(t,\tau)$ read
\begin{eqnarray*}
\phi_{\rm L\,1}(t,\tau)&=& \frac{\partial \bar{W}_{\rm L}}{\partial I_{1}}
\Bigl (T(t), I_{1}({\bf C}(t,\tau)),I_{2}({\bf C}(t,\tau))\Bigr )
\nonumber\\
&& +I_{1}({\bf C}(t,\tau))\frac{\partial \bar{W}_{\rm L}}{\partial I_{2}}
\Bigl (T(t), I_{1}({\bf C}(t,\tau)),I_{2}({\bf C}(t,\tau))\Bigr ),
\nonumber\\
\phi_{\rm L\,2}(t,\tau) &=& -\frac{\partial \bar{W}_{\rm L}}{\partial I_{2}}
\Bigl (T(t), I_{1}({\bf C}(t,\tau)),I_{2}({\bf C}(t,\tau))\Bigr ).
\end{eqnarray*}
It follows from Eqs. (31) and (35) that
\begin{eqnarray}
\sum_{k=1}^{2} U_{k}^{\circ}(t)\frac{dI_{k}}{dt}({\bf C}^{\circ}(t))
&=& 2\Bigl \{ (1-\mu_{\rm L}-\mu_{\rm H}){\bf \Lambda}_{\rm B}^{\circ}(t)
+\mu_{\rm H}{\bf \Lambda}_{\rm H}^{\circ}(t)
\nonumber\\
&&+\mu_{\rm L}{\bf \Lambda}_{\rm L}^{\circ}(t)\int_{0}^{\infty}
\exp \Bigl [-\int_{0}^{t} \Gamma(s,\omega)ds\Bigr ]
p(\omega)d\omega \Bigr \}:{\bf D}(t),
\end{eqnarray}
where
\begin{equation}
{\bf \Lambda}_{\rm \beta}^{\circ}(t)
= \phi_{\beta\,1}^{\circ}(t){\bf F}^{\circ}(t)
+\phi_{\beta\,2}^{\circ}(t)\Bigl [ {\bf F}^{\circ}(t)\Bigr ]^{2}
\qquad
(\beta={\rm B,L,H}),
\end{equation}
and the functions $\phi_{\beta\,k}^{\circ}(t)$ are given by
\begin{eqnarray*}
\phi_{\rm B\,1}^{\circ}(t)&=& \frac{\partial \bar{W}_{\rm B}}{\partial I_{1}}
\Bigl (T(t), I_{1}({\bf C}^{\circ}(t)),I_{2}({\bf C}^{\circ}(t))\Bigr )
\nonumber\\
&& +I_{1}({\bf C}^{\circ}(t))
\frac{\partial \bar{W}_{\rm B}}{\partial I_{2}}
\Bigl (T(t), I_{1}({\bf C}^{\circ}(t)),I_{2}({\bf C}^{\circ}(t))\Bigr ),
\nonumber\\
\phi_{\rm L\,1}^{\circ}(t)&=& \frac{\partial \bar{W}_{\rm L}}{\partial I_{1}}
\Bigl (T(t), I_{1}({\bf C}^{\circ}(t)),I_{2}({\bf C}^{\circ}(t))\Bigr )
\nonumber\\
&& +I_{1}({\bf C}^{\circ}(t))
\frac{\partial \bar{W}_{\rm L}}{\partial I_{2}}
\Bigl (T(t), I_{1}({\bf C}^{\circ}(t)),I_{2}({\bf C}^{\circ}(t))\Bigr ),
\nonumber\\
\phi_{\rm H\,1}^{\circ}(t)&=& \frac{\partial \bar{W}_{\rm H}}{\partial I_{1}}
\Bigl ({\bf a}(t),T(t), I_{1}({\bf C}^{\circ}(t)),I_{2}({\bf C}^{\circ}(t))\Bigr )
\nonumber\\
&& +I_{1}({\bf C}^{\circ}(t))\frac{\partial \bar{W}_{\rm H}}{\partial I_{2}}
\Bigl ({\bf a}(t),T(t), I_{1}({\bf C}^{\circ}(t)),I_{2}({\bf C}^{\circ}(t))\Bigr ),
\nonumber\\
\phi_{\rm B\,2}^{\circ}(t)&=& - \frac{\partial \bar{W}_{\rm B}}{\partial I_{2}}
\Bigl (T(t), I_{1}({\bf C}^{\circ}(t)),I_{2}({\bf C}^{\circ}(t))\Bigr ),
\nonumber\\
\phi_{\rm L\,2}^{\circ}(t)&=& - \frac{\partial \bar{W}_{\rm L}}{\partial I_{2}}
\Bigl (T(t), I_{1}({\bf C}^{\circ}(t)),I_{2}({\bf C}^{\circ}(t))\Bigr ),
\nonumber\\
\phi_{\rm H\,2}^{\circ}(t)&=& - \frac{\partial \bar{W}_{\rm H}}{\partial I_{2}}
\Bigl ({\bf a}(t),T(t), I_{1}({\bf C}^{\circ}(t)),I_{2}({\bf C}^{\circ}(t))\Bigr ).
\end{eqnarray*}
Combining Eqs. (30), (36) and (38), we arrive at the formula
\begin{equation}
\frac{dW}{dt}(t) = J(t)\frac{dT}{dt}(t)+{\bf A}(t)\cdot\frac{d{\bf a}}{dt}(t)
+2{\bf H}(t):{\bf D}(t)-Y(t)
\end{equation}
with
\begin{eqnarray}
{\bf H}(t) &=& (1-\mu_{\rm L}-\mu_{\rm H}){\bf \Lambda}_{\rm B}^{\circ}(t)
+\mu_{\rm H}{\bf \Lambda}_{\rm H}^{\circ}(t)
\nonumber\\
&&+\mu_{\rm L}\Bigl \{ {\bf \Lambda}_{\rm L}^{\circ}(t)\int_{0}^{\infty}
\exp \Bigl [-\int_{0}^{t} \Gamma(s,\omega)ds\Bigr ] p(\omega)d\omega
+{\bf \Lambda}_{\rm L}(t)\Bigr \}.
\end{eqnarray}

\section{Constitutive equations}

For an incompressible medium, the Clausius--Duhem inequality reads \cite{CG67}
\begin{equation}
T \frac{dQ}{dt}=-S\frac{dT}{dt}-\frac{d\Psi}{dt}
+\frac{1}{\rho}\Bigl ( {\bf \Sigma}^{\prime}:{\bf D}-\frac{1}{T}{\bf q}\cdot
{\bf \nabla} T \Bigr) \geq 0,
\end{equation}
where $\rho$ is mass density,
${\bf \Sigma}^{\prime}$ is the deviatoric component
of the Cauchy stress tensor ${\bf \Sigma}$,
${\bf q}$ is the heat flux vector,
$S$ is the entropy per unit mass,
and $Q$ is the entropy production.

It follows from Eqs. (1), (2) and (40) that
\begin{equation}
\frac{d\Psi}{dt}=-\Bigl ( S^{\circ}+c\ln\frac{T}{T^{\circ}}-J\Bigr )\frac{dT}{dt}
+{\bf A}\cdot\frac{d{\bf a}}{dt}+2{\bf H}:{\bf D}-Y.
\end{equation}
Equations (42) and (43) imply that
\begin{eqnarray}
T \frac{dQ}{dt}&=& -\Bigl ( S-S^{\circ}-c\ln\frac{T}{T^{\circ}}+J\Bigr )\frac{dT}{dt}
+\frac{1}{\rho}\Bigl ( {\bf \Sigma}^{\prime}-2\rho {\bf H}\Bigr ):{\bf D}
\nonumber\\
&& +{\bf A}\cdot\frac{d{\bf a}}{dt}+Y-\frac{1}{\rho T}{\bf q}\cdot
{\bf \nabla} T \geq 0.
\end{eqnarray}
Because inequality (44) is to be satisfied for an arbitrary
loading program, the expressions in brackets should vanish.
This results in the formula for the entropy per unit mass
\begin{equation}
S(t)=S^{\circ}+c\ln\frac{T(t)}{T^{\circ}}-J(t)
\end{equation}
and the constitutive equation
\begin{eqnarray}
{\bf \Sigma}(t) &=& -P(t){\bf I}+2\rho \Bigr \{
(1-\mu_{\rm L}-\mu_{\rm H}){\bf \Lambda}_{\rm B}^{\circ}(t)
+\mu_{\rm H}{\bf \Lambda}_{\rm H}^{\circ}(t)
\nonumber\\
&&+\mu_{\rm L}\Bigl [ {\bf \Lambda}_{\rm L}^{\circ}(t)\int_{0}^{\infty}
\exp \Bigl (-\int_{0}^{t} \Gamma(s,\omega)ds\Bigr ) p(\omega)d\omega
+{\bf \Lambda}_{\rm L}(t) \Bigr ]\Bigr \},
\end{eqnarray}
where $P$ is pressure.
To ensure that the rate of entropy production is non-negative,
it is natural to suppose that the derivative of the vector
of internal variables ${\bf a}$ with respect to time is proportional
to the vector ${\bf A}$, which implies the evolution equation
\begin{equation}
\frac{d{\bf a}}{dt}=\eta {\bf A},
\end{equation}
where $\eta\geq 0$ is a function to be determined.
We assume that $\eta$ depends on the vector ${\bf a}$,
the rate-of-strain tensor ${\bf D}$,
and the Cauchy deformation tensor ${\bf C}^{\circ}$,
\[ \eta=\eta({\bf a},{\bf D}, {\bf C}^{\circ}). \]
To distinguish explicitly the viscoelastic behavior of filled rubber and
damage of filler clusters, it is postulated that the coefficient $\eta$
vanishes for a time-independent loading process
\begin{equation}
\eta({\bf a},{\bf D}, {\bf C}^{\circ})\Bigl |_{{\bf D}={\bf 0}}=0.
\end{equation}
By analogy with Eq. (47), we adopt the Fourier law for the heat flux vector ${\bf q}$,
\begin{equation}
{\bf q}=-\kappa {\bf \nabla}T
\end{equation}
where $\kappa>0$ is a coefficient of thermal diffusivity.

Substituting expressions (45) to (47) and (49) into Eq. (44)
and bearing in mind formula (31) for the functional $Y$,
we find that for any loading process, the Clausius--Duhem inequality (44) is satisfied.

\section{Uniaxial extension of a specimen}

We now apply constitutive equation (46) to determine stresses in a bar
at uniaxial extension.
Points of the bar refer to Cartesian coordinates $\{ X_{i} \}$
in the reference (stress-free) state and to Cartesian coordinates $\{ x_{i} \}$
in the deformed state, $(i=1,2,3)$.
Tension of an incompressible medium is described by the formulas
\begin{equation}
x_{1}=k(t)X_{1},
\qquad
x_{2}=k^{-\frac{1}{2}}(t) X_{2},
\qquad
x_{3}=k^{-\frac{1}{2}}(t) X_{3},
\end{equation}
where $k=k(t)$ is the extension ratio.
It follows from Eq. (50) that the deformation gradient, ${\bf \nabla}^{\circ}{\bf R}(t)$,
and the relative deformation gradient, ${\bf \nabla}(\tau){\bf R}(t)$, read
\begin{eqnarray*}
{\bf \nabla}^{\circ}{\bf R}(t) &=& k(t){\bf e}_{1}{\bf e}_{1}
+\biggl (\frac{1}{k(t)}\biggr )^{\frac{1}{2}}
({\bf e}_{2}{\bf e}_{2}+{\bf e}_{3}{\bf e}_{3}),
\nonumber\\
{\bf \nabla}(\tau){\bf R}(t) &=& \frac{k(t)}{k(\tau)}{\bf e}_{1}{\bf e}_{1}
+\biggl (\frac{k(\tau)}{k(t)}\biggr )^{\frac{1}{2}}
({\bf e}_{2}{\bf e}_{2}+{\bf e}_{3}{\bf e}_{3}),
\end{eqnarray*}
where ${\bf e}_{i}$ are base vectors of the frame $\{ X_{i} \}$.
The Finger tensor, ${\bf F}^{\circ}(t)$, and the relative Finger tensor,
${\bf F}(t,\tau)$, are given by
\begin{eqnarray}
{\bf F}^{\circ}(t) &=& k^{2}(t){\bf e}_{1}{\bf e}_{1}+\frac{1}{k(t)}
({\bf e}_{2}{\bf e}_{2}+{\bf e}_{3}{\bf e}_{3}),
\nonumber\\
{\bf F}(t,\tau) &=& \biggl (\frac{k(t)}{k(\tau)}\biggr )^{2}
{\bf e}_{1}{\bf e}_{1}+\frac{k(\tau)}{k(t)}
({\bf e}_{2}{\bf e}_{2}+{\bf e}_{3}{\bf e}_{3}).
\end{eqnarray}
Substituting Eqs. (37), (39) and (51) into Eq. (46), we find the non-zero components
\begin{eqnarray}
\Sigma_{1}(t) &=& -P(t)+ 2\rho \biggl \{ (1-\mu_{\rm L}-\mu_{\rm H})
\Bigl [ \phi_{\rm B\, 1}^{\circ}(t)+\phi_{\rm B\, 2}^{\circ}(t)k^{2}(t)\Bigr ]
+\mu_{\rm H}\Bigl [ \phi_{\rm H\, 1}^{\circ}(t)+\phi_{\rm H\, 2}^{\circ}(t)k^{2}(t)\Bigr ]
\nonumber\\
&&+\mu_{\rm L}\Bigl [ \phi_{\rm L\, 1}^{\circ}(t)+\phi_{\rm L\, 2}^{\circ}(t)k^{2}(t)\Bigr ]
\int_{0}^{\infty} \exp \Bigl [-\int_{0}^{t} \Gamma(s,\omega)ds\Bigr ]
p(\omega)d\omega \biggr \} k^{2}(t)
\nonumber\\
&& +2\rho \mu_{\rm L}\int_{0}^{\infty} p(\omega) d\omega \int_{0}^{t} \biggl [
\phi_{\rm L\, 1}(t,\tau)\biggl (\frac{k(t)}{k(\tau)}\biggr )^{2}
+\phi_{\rm L\, 2}(t,\tau)\biggl (\frac{k(t)}{k(\tau)}\biggr )^{4}\biggr ]
\nonumber\\
&&\times
\Gamma(\tau,\omega)\exp \Bigl [-\int_{\tau}^{t} \Gamma(s,\omega)ds\Bigr ]d\tau,
\nonumber\\
\Sigma_{2}(t) &=& -P(t)+ 2\rho \biggl \{ (1-\mu_{\rm L}-\mu_{\rm H})
\Bigl [ \phi_{\rm B\, 1}^{\circ}(t)+\phi_{\rm B\, 2}^{\circ}(t)k^{-1}(t)\Bigr ]
+\mu_{\rm H}\Bigl [ \phi_{\rm H\, 1}^{\circ}(t)+\phi_{\rm H\, 2}^{\circ}(t)k^{-1}(t)\Bigr ]
\nonumber\\
&&+\mu_{\rm L}\Bigl [ \phi_{\rm L\, 1}^{\circ}(t)+\phi_{\rm L\, 2}^{\circ}(t)k^{-1}(t)\Bigr ]
\int_{0}^{\infty} \exp \Bigl [-\int_{0}^{t} \Gamma(s,\omega)ds\Bigr ]
p(\omega)d\omega \biggr \} k^{-1}(t)
\nonumber\\
&& +2\rho \mu_{\rm L}\int_{0}^{\infty} p(\omega) d\omega \int_{0}^{t} \biggl [
\phi_{\rm L\, 1}(t,\tau)\biggl (\frac{k(\tau)}{k(t)}\biggr )
+\phi_{\rm L\, 2}(t,\tau)\biggl (\frac{k(\tau)}{k(t)}\biggr )^{2}\biggr ]
\nonumber\\
&&\times
\Gamma(\tau,\omega)\exp \Bigl [-\int_{\tau}^{t} \Gamma(s,\omega)ds\Bigr ]d\tau
\end{eqnarray}
of the Cauchy stress tensor
\[ {\bf \Sigma}=\Sigma_{1}{\bf e}_{1}{\bf e}_{1}
+\Sigma_{2} ({\bf e}_{2}{\bf e}_{2}+{\bf e}_{3}{\bf e}_{3}). \]
Excluding the unknown pressure $P(t)$ from Eq. (52) and the boundary condition
\[ \Sigma_{2}(t)=0, \]
we arrive at the formula for the longitudinal stress
\begin{eqnarray}
\Sigma_{1}(t) &=& 2\rho \biggl \{ (1-\mu_{\rm L}-\mu_{\rm H})
\phi_{\rm B\, 1}^{\circ}(t)+\mu_{\rm H}\phi_{\rm H\, 1}^{\circ}(t)
\nonumber\\
&&+\mu_{\rm L}\phi_{\rm L\, 1}^{\circ}(t)
\int_{0}^{\infty} \exp \Bigl [-\int_{0}^{t} \Gamma(s,\omega)ds\Bigr ]
p(\omega)d\omega
\nonumber\\
&&+\Bigl [(1-\mu_{\rm L}-\mu_{\rm H})
\phi_{\rm B\, 2}^{\circ}(t)+\mu_{\rm H}\phi_{\rm H\, 2}^{\circ}(t)
\nonumber\\
&& +\mu_{\rm L}\phi_{\rm L\, 2}^{\circ}(t)
\int_{0}^{\infty} \exp \Bigl (-\int_{0}^{t} \Gamma(s,\omega)ds\Bigr )
p(\omega)d\omega \Bigr ]
\nonumber\\
&&\times \Bigl [k^{2}(t)+k^{-1}(t)\Bigr ]
\biggr \} \Bigl [k^{2}(t)-k^{-1}(t)\Bigr ]
\nonumber\\
&& +2\rho \mu_{\rm L}\int_{0}^{\infty} p(\omega) d\omega \int_{0}^{t}
\biggl \{ \phi_{\rm L\, 1}(t,\tau)
+\phi_{\rm L\, 2}(t,\tau)\biggl [ \biggl (\frac{k(t)}{k(\tau)}\biggr )^{2}
+\frac{k(\tau)}{k(t)}\biggr ]\biggr \}
\nonumber\\
&&\times
\biggl [ \biggl (\frac{k(t)}{k(\tau)}\biggr )^{2}-\frac{k(\tau)}{k(t)}\biggr ]
\Gamma(\tau,\omega)\exp \Bigl [-\int_{\tau}^{t} \Gamma(s,\omega)ds\Bigr ]d\tau.
\end{eqnarray}
Equation (53) describes the mechanical response of a specimen for an arbitrary
loading program.
To compare results of numerical simulation with experimental data, we focus
on the relaxation test with
\begin{equation}
k(t)=\left \{ \begin{array}{lll}
1, && t<0,\\
\lambda, && t>0,
\end{array}
\right .
\end{equation}
where $\lambda >1$ is a constant.
Substitution of Eq. (54) into Eq. (53) results in
\begin{equation}
\Sigma_{1}(t)=\Bigl [ \sigma_{\rm B}+\sigma_{\rm L}+\sigma_{\rm H}-\sigma(t)\Bigr ]
(\lambda^{2}-\lambda^{-1}),
\end{equation}
where the parameters
\begin{eqnarray*}
\sigma_{\rm B} &=& 2\rho (1-\mu_{\rm L}-\mu_{\rm H}) \Bigl [ \phi_{\rm B\,1}^{\circ}
+ \phi_{\rm B\,2}^{\circ}(\lambda^{2}+\lambda^{-1})\Bigr ],
\nonumber\\
\sigma_{\rm L} &=& 2\rho \mu_{\rm L} \Bigl [ \phi_{\rm L\,1}^{\circ}
+ \phi_{\rm L\,2}^{\circ}(\lambda^{2}+\lambda^{-1})\Bigr ],
\nonumber\\
\sigma_{\rm H} &=& 2\rho \mu_{\rm H} \Bigl [ \phi_{\rm H\,1}^{\circ}
+ \phi_{\rm H\,2}^{\circ}(\lambda^{2}+\lambda^{-1})\Bigr ]
\end{eqnarray*}
are independent of time and
\begin{equation}
\sigma(t) = \sigma_{\rm L}\int_{0}^{\infty}\biggl [ 1
-\exp\biggl (-\int_{0}^{t} \Gamma(s,\omega)ds \biggr )\biggr ] p(\omega)d\omega.
\end{equation}
It follows from Eqs. (13), (55) and (56) that the dimensionless ratio
\[ R(t)=\frac{\Sigma_{1}(t)}{\Sigma_{1}(0)} \]
is given by
\begin{equation}
R(t)= 1-A\int_{0}^{\infty}\biggl [ 1
-\exp\biggl (-\Gamma_{\ast} \exp(-\omega)t \biggr )\biggr ]p(\omega)d\omega,
\end{equation}
where
\begin{equation}
A=\frac{\sigma_{\rm L}}{\sigma_{\rm B}+\sigma_{\rm L}+\sigma_{\rm H}}.
\end{equation}
Equations (12) and (57) are determined by 4 adjustable parameters:
the constant $A$ is the ratio of the relaxing stress to the
total stress at the beginning of the test,
$\Gamma_{\ast}$ characterizes the average rate of relaxation,
$\Omega$ and $\Sigma$ in Eq. (12) describe the distribution
of relaxation times.

Our objective now is to find the constants $A$, $\Gamma_{\ast}$, $\Omega$ and $\Sigma$
by fitting experimental data in relaxation tests with various
elongation ratios, $\lambda$, for unfilled and CB filled rubbers.

\section{Experimental}

To assess the time-dependent response of particle-reinforced elastomers,
a series of uniaxial relaxation tests were carried out at a constant
temperature.
Dumbbell specimens were provided by TARRC laboratories
(Hertford, UK) and were used as received.

The first compound on the base of natural rubber (EDS--19)
contains 1 phr (parts per hundred parts of rubber) of carbon black
for UV protection.
The filler content is too low to affect the mechanical response
significantly, and we refer to this compound as an unfilled rubber.

The other compound (EDS--16) contains 45 phr of carbon black
and this compound is treated as a filled rubber.
Before the vulcanization process, a sheet of rubber from which
the specimens were cut out was milled in a rolling mill to destroy
aggregates of filler particles.
This procedure leads to a weak anisotropy of the material.
The dumbbell specimens cut in the direction of milling are
referred to as R1, whereas the samples cut in
the orthogonal direction are abbreviated as R2.

A detailed chemical formulation of compounds EDS--16 and EDS--19
is presented in Table 1.

The relaxation tests were performed at room temperature by using a testing
machine designed at the Institute of Physics (Vienna, Austria).
A specimen was loaded with the strain rate $5.0 \cdot 10^{-3}$ (s$^{-1}$)
up to a given elongation ratio $\lambda$, which was preserved constant
during the relaxation test (1 hour).
To measure the longitudinal strain, two reflection lines were drawn
in the central part of each specimen before loading
(with the distance 7 mm between them).
Changes in the distance between these lines were controlled by
a video-extensometer (which ensured the accuracy of about 1 \%).

The tensile force was measured by using a standard loading cell.
The nominal stress was determined as the ratio of the axial force
to the cross-sectional area of a specimen (1 mm $\times$ 4 mm)
in the stress-free state.

The relaxation process was studied at three elongation ratios
($\lambda=1.2$, $\lambda=1.4$ and $\lambda=1.8$) for samples made of
the unfilled rubber,
and at four elongation ratios ($\lambda=1.2$, $\lambda=1.4$, $\lambda=1.8$
and $\lambda=2.0$) for specimens made of the filled rubbers R1 and R2.
None of the specimens was subjected to heat treatment
or mechanical pre-loading prior to testing.

\section{Comparison with experimental data}

We begin with matching observations in relaxation tests for unfilled rubber.
First, we fit experimental data in the test with the minimum elongation
ratio $\lambda_{\min}=1.2$.
Because the parameters $\Gamma_{\ast}$ and $\Omega$ appear to be mutually
dependent (an increase in $\Omega$ leads to the growth of $\Gamma_{\ast}$),
we set $\Gamma_{\ast}=1.0$ (s$^{-1}$) in the approximation
of experimental data at $\lambda_{\min}$.
Given $A$, the quantities $\Omega$ and $\Sigma$ are found by the steepest-descent
method.
The coefficient $A$ is determined by the least-squares technique.

Afterwards, we match observations for the unfilled rubber
at the elongation ratios $\lambda=1.4$ and $\lambda=1.8$
by using the parameters $\Omega$ and $\Sigma$ which were found
in fitting data at $\lambda_{\min}$.
The relaxation rate $\Gamma_{\ast}$ is determined by the steepest-descent
procedure, and the coefficient $A$ is found by the least-squares method.
Figure 1 demonstrates fair agreement between observations and results
of numerical simulation.
It is worth noting that relaxation curves at different elongation ratios, $\lambda$,
noticeably differ from each other, despite some scatter in the experimental data.

The average relaxation time, $\tau_{\ast}=\Gamma_{\ast}^{-1}$, and the
dimensionless coefficient $A$ are plotted in Figures 2 and 3 versus the
first invariant of the Cauchy deformation tensor, $I_{1}({\bf C}^{\circ})$.
The effect of straining on $\tau_{\ast}$ and $A$ is adequately described
by the phenomenological equations
\begin{equation}
\tau_{\ast}=\tau_{0}+\tau_{1}(I_{1}-3),
\qquad
A=A_{0}+A_{1}(I_{1}-3),
\end{equation}
where the constants $\tau_{k}$ ($\tau_{0}\geq 0$) and $A_{k}$ ($k=0,1$) are determined
by the least-squares method.

We proceed with matching experimental data for two grades (R1
and R2) of filled rubber by applying the same procedure of fitting.
Observations and results of numerical analysis are depicted in Figures 4 and 5
which demonstrate rather small scatter of experimental data.
The parameters $\tau_{\ast}$ and $A$ are plotted in Figures 6 and 7
together with their approximations by Eq. (59).

To compare our findings with other observations,
we apply the same procedure of fitting to data for a filled rubber
obtained in tensile relaxation tests by Haupt and Sedlan \cite{HS01}
(the samples used in their study are abbreviated as R3).
For a description of specimens and the experimental procedure, the reader
is referred to the paper \cite{HS01}.
Figure 8 demonstrates good agreement between observations and results of
numerical analysis.
The experimental constants $\tau_{\ast}$ and $A$ are plotted in Figures 9 and
10 together with predictions of Eq. (59).

Adjustable parameters for various grades of unfilled and CB filled
rubbers are listed in Table 2.

\section{Discussion}

Experimental data depicted in Figures 1, 4, and 5 reveal that the response of
the rubbers under consideration (both carbon black filled and unfilled)
in the standard relaxation tests strongly depends on time.
The relative decrease in stresses after relaxation for $t_{0}=1$ hour
at the minimum elongation ratio $\lambda_{\min}=1.2$,
\[ \zeta(t_{0})=\frac{\Sigma_{1}(0)-\Sigma_{1}(t_{0})}{\Sigma_{1}(0)} \]
equals 6 \% for the unfilled rubber and varies from 13 to 15 \%
for the CB filled rubbers.
The quantity $\zeta(t_{0})$ for the filled rubbers (R1 and R2)
noticeably exceeds (about by twice) that for the unfilled rubber.
The fact that the presence of filler accelerates relaxation in rubbers
may be explained by an increase in the number of regions with low concentration
of cross-links during the mixing process in a particle-reinforced elastomer
compared to that for an unfilled medium.
However, because the amounts $\zeta(t_{0})$ have the same order of
magnitude for unfilled and filled rubbers, we suppose that
the viscoelastic behavior of elastomers cannot be described
exclusively by means of the mechanically induced destruction of CB clusters
and de-wetting of macromolecules from the filler particles \cite{AH96}.

The relaxation spectrum (which is characterized by the distribution function, $p(\omega)$,
for strands with various energies for breakage, $\omega$) is independent of the
elongation ratio, $\lambda$, at least, in the region of moderate deformations, $\lambda\in (1,2)$.
The distribution function, $p(\omega)$, is determined by two adjustable parameters,
$\Omega$ and $\Sigma$, which accept similar values for the three grades
of filled rubber (R1, R2, and R3).
Their values, however, substantially differ from those for the unfilled rubber.
Because the ratio $\xi=\Sigma/\Omega$ lies in the interval between 0.5 and 0.6 for all specimens,
we may associate $\Omega$ with the average
energy and $\Sigma$ with the standard deviation of energy for breakage.
Table 2 demonstrates that the average energy for breakage
of strands, $\Omega$, for the unfilled rubber exceeds by twice (approximately)
the average energy for breakage for the filled rubbers, whereas the ratio $\xi$
(that characterizes the width of the Gaussian distribution)
is practically independent of the presence of filler.
This observation means that reinforcement of rubber by CB particles
results in an increase in the level of inhomogeneity of the compound
and in the growth of the number of regions with low concentration of junctions
where reformation of strands occurs.

Figures 2, 6, and 9 evidence that the characteristic time of relaxation, $\tau_{\ast}$,
increases with the growth of longitudinal elongation $\lambda$.
This conclusion contradicts the strain--time superposition principle
conventionally employed in the analysis of the viscoelastic response of polymeric glasses,
see, e.g., \cite{KE87,LK92}.
According to the free-volume concept, mechanical loading
results in an increase in the free volume, which, in turn, causes an acceleration of
the reformation process for mobile units (cooperatively rearranging regions).
The free-volume theory correctly describes the time-dependent behavior of
glassy polymers, where mechanically induced changes in the specific volume are substantial.
The applicability of that approach to rubbery polymers appears to be
questionable, because of the incompressibility of rubbers in the region of moderate
strains (up to $\lambda=3$ for uniaxial tension).

The growth in the characteristic time for rearrangement of strands, $\tau_{\ast}$,
with $\lambda$ may be explained by the strain-induced crystallization of natural
rubber.
Uniaxial extension of a specimen at the macro-level causes stretching
of strands at the micro-level.
This results in the creation of nuclei of crystallites (regions of strong order
and high density) in the vicinity of junctions, which slow down breakage
and reformation of chains.
With reference to this picture, one can hypothesize that the presence of filler
particles (which may be thought of as obstacles for the formation of
micro-crystallites) should result in a less pronounced increase of
the average time for reformation of strands, $\tau_{\ast}$, with the growth
of macro-strains.
This conclusion is confirmed by the data listed in Table 2 which
reveal that the parameter $\tau_{1}$ in Eq. (59) (the ``rate of growth"
for the characteristic time of relaxation with
the measure of deformation, $I_{1}-3$) for the unfilled rubber exceeds
that for the filled rubbers by several times.

An increase in the average relaxation time, $\tau_{\ast}$,
with stretching of a specimen is correctly predicted by Eq. (59)
for the unfilled rubber and for the filled rubber R1.
For the other two grades of particle-reinforced rubber (R2 and R3),
$\tau_{\ast}$ falls down rapidly in the interval
between $\lambda=1.4$ and $\lambda=1.8$ and proceeds to increase
for larger values of the elongation ratio.
This decrease in $\tau_{\ast}$ (or the corresponding increase in
the rate of breakage $\Gamma_{\ast}$) may be explained if we assume
that links between filler particles and long chains
are strong enough in order to prevent thermally induced detachment
of strands from the particles at relatively small strains.
This implies that in the region of relatively small elongation ratios
($\lambda \leq 1.4$) particles of filler may be treated as
permanent cross-links between chains.
When the macro-deformation of a specimen reaches a threshold
value for the load-bearing capacity of these links,
chains begin to slip from the particles, which is reflected by
the model as a noticeable increase in the rate of breakage $\Gamma_{\ast}$.
The parameter $\tau_{1}$ (that describes the effect of stretching
on the characteristic time of relaxation) in the region of
low elongation ratios ($0\leq \lambda\leq 1.4$) substantially exceeds that
in the domain of large stretches ($1.8\leq\lambda\leq 2.0$).
This may be explained by the interaction of two physical phenomena in
the region of large deformations: (i) the growth of $\lambda$ results
in slowing down of the reformation rate (which is in common for
regions of low and high strains), and (ii) an increase in the
concentration of temporary cross-links (when the stretch intensity
exceeds the load-bearing capacity of links between macromolecules
and filler particles) accelerates the reformation process
(which is a specific feature for the domain of relatively large
strains).

It is worth noting that the parameter $\tau_{1}$ for the filled rubber R2
(samples cut out in the direction orthogonal to the direction of milling)
noticeably exceeds that for the filled rubber R1 (specimens cut in the
direction of milling).
This may be explained by the mechanically induced crystallization of
specimens before their vulcanization.
Chains with the end-to-end vectors directed along the direction of loading
in the relaxation tests are partially crystallized during milling,
which results in an essential decrease in their rate of breakage
and reformation.
Because these chains bear the main load under tension for the specimens R1,
the influence of macro-deformation on the breakage process becomes
relatively weak (small values of $\tau_{1}$).
On the contrary, partial crystallization of chains with the end-to-end vectors
directed perpendicular to the direction of stretching for specimens R2
does not affect their rates of breakage (because micro-strains in these
chains are relatively small), whereas the chains directed along
the stretching direction are practically not crystallized, and their
response is similar to that for the unfilled rubber (large values of
$\tau_{1}$).

Figures 3 and 7 demonstrate that $A$ increases with
the longitudinal strain, and this dependence can be correctly described
by the phenomenological equation (59).
According to Table 2, the value of $A$ in the stress-free state, $A_{0}$,
equals approximately 0.2 for all compounds under consideration.
Bearing in mind that $A$ is the ratio of the stress in regions with
low concentration of junctions to the total stress in a specimen,
see Eq. (58), and presuming (as a first approximation) that the stress
is linearly proportional to the mechanical energy,
we conclude that the mechanical energy for RLCJs equals about 0.2
of the total strain energy for rubber.
Finally, assuming that the energy stored in a domain is proportional to
its volume, we arrive at the rough estimate for the volume occupied by regions
where relaxation of stresses occurs: about 20 \% of the total volume of
a specimen (Table 2 shows that this value weakly depends on the content
of filler).

Equation (59) implies that an increase in the parameter $A$ with
the elongation ratio $\lambda$ means that the stress, $\sigma_{\rm L}$,
in RLCJs grows more rapidly then the stresses in the bulk material,
$\sigma_{\rm B}$, and in the regions with high concentration of junctions,
$\sigma_{\rm H}$.
This conclusion is fulfilled both for the unfilled rubber and for the filled
rubbers R1 and R2 (which were subjected to milling during the preparation
process).
On the contrary, Figure 10 demonstrates that at relatively small elongation
ratios, $\lambda\leq 1.6$, the parameter $A$ decreases, whereas at large
stretches, $\lambda\geq 1.7$, it linearly increases with the first principal
invariant, $I_{1}$, of the Cauchy deformation tensor, ${\bf C}^{\circ}$.
This non-monotonicity in the dependence of $A$ on strains for specimens R3
may be explained within the concept of occluded rubber \cite{WRC93}.
According to that theory, aggregation of isolated clusters of particles
into a network results in the separation of some regions of the host material
(occluded domains) from the bulk elastomer.
When the secondary network of filler is relatively rigid,
it does not transmit stresses to the occluded domains.
As a consequence, at small strains, occluded rubber is not deformed,
and its mechanical energy vanishes.
With the growth of strains, the network of filler breaks, and some regions of the
occluded rubber are released from the constrains.
This results in an increase in the total strain energy and stresses
in the bulk material.
With reference to Eq. (58), this conclusion implies that in the region of
relatively small elongation ratios ($\lambda\leq 1.6$) for rubber R3,
the sum $\sigma_{\rm B}+\sigma_{\rm H}$ grows faster then the stress, $\sigma_{\rm L}$,
in the domains with low concentration of junctions.
When the elongation ratio $\lambda$ reaches its threshold value
($\lambda\approx 1.7$), all regions of occluded rubber are released from
restrains, and the growth of strains implies a more rapid increase
in the stress $\sigma_{\rm L}$ then in the stresses in the host elastomer
and in the regions with high concentration of junctions.
The latter is observed as the growth in the parameter $A$.
The difference in the behavior of the function $A(I_{1})$ for the unfilled rubber
and for the CB filled rubbers R1 and R2, on the one hand, and for the rubber R3, on the other
hand, may be explained by the absence of occluded regions for the first three compounds
(the regions of occluded rubber are supposed to be destroyed during milling)
and the presence of these regions in the samples R3 with no preliminary mechanical
treatment.

It follows from Figures 3 and 7 that for any elongation ratio $\lambda\geq 0$,
the quantity $A$ is maximal for the unfilled rubber, and the value of $A$
for specimens cut in the direction orthogonal to the direction of milling (R2)
is higher then that for specimens cut in the direction of milling (R1).
The fact that the ratio $A$ for the unfilled rubber exceeds that for the filled
rubbers (R1 and R2) follows from Eq. (58), provided that the stress
in the regions with high concentration of junctions exceeds that for the bulk material.
The observation that the parameter $A$ for specimens R2 exceeds that for the
specimens R1 confirms our hypothesis about the effect of mechanically induced
crystallization on the response of filled rubbers, because $\sigma_{\rm B}$
increases when macromolecules whose end-to-end vectors are directed along
the axis of extension are partially crystallized.

\section{Concluding remarks}

Constitutive equations have been derived for the isothermal viscoelastic
behavior of unfilled and filled elastomers at finite strains.
The model is based on the concept of temporary networks that treats
a rubbery polymer as a network of long chains connected to junctions
(chemical and physical cross-links and entanglements).
A filled rubber is thought of as a strongly non-homogeneous medium
(with the characteristic length of inhomogeneity of about 100 nm)
which consists of an elastomeric matrix with a normal concentration
of cross-links, wherein regions with low and high concentrations of
cross-links are distributed.
In the domains with low concentration of cross-links,
the average number of junctions between polymeric chains (per unit mass)
is less then that in the bulk material because of local inhomogeneities
in the distribution of a cross-linker during the mixing procedure.
In the domains with high concentration of cross-links,
the number of junctions between long chains exceeds that for the host elastomer,
because particles of filler and their clusters play the role of extra cross-links.

In the bulk material and in the regions with high concentration of cross-links,
intermolecular forces prevent slippage of chains from junctions (which are
treated as permanent cross-links).
In the domains with low concentration of cross-links, active strands (whose ends
are linked to separate junctions) are assumed to slip from the temporary junctions
as they are thermally agitated.
Breakage of an active strand (its transition into a dangling strand)
occurs at a random instant when one of its ends detaches from a junction.
A dangling strand is reformed (it merges with the network) at a random time
when its free end captures some junction in its vicinity.
The stress in a dangling strand is assumed to totally relax before this strand
is attached to the network, which implies that the natural (stress-free)
state of any chain coincides with the deformed state of the network at the instant of
their merging.
Unlike conventional theories of transient networks, we postulate that different strands
in RLCJs have different energies for breakage and describe these energies by
the quasi-Gaussian distribution function (12) with two adjustable parameters, $\Omega$ and $\Sigma$.

Constitutive equations for a filled rubber are developed by using the laws of
thermodynamics.
These equations are applied to determine stresses at uniaxial extension of a bar.
The time-dependent response of a specimen in the standard relaxation
test is characterized by four adjustable parameters which are found by
fitting observations.
To determine these constants, a series of tests have been carried out on
specimens of unfilled and CB filled natural rubber.
Fair agreement is demonstrated between results of numerical simulation
and observations in relaxation tests (with the duration of $t_{0}=1$ hour)
at various elongation ratios, $\lambda$, in the region of moderate finite
deformations, $\lambda\in (1,2)$.

The following conclusions are drawn:
\begin{enumerate}
\item
The unfilled and particle reinforced rubbers demonstrate a pronounced
viscoelastic response.
The relative decrease in stresses, $\zeta(t_{0})$,
for the filled rubbers exceeds that for the unfilled rubber by twice.
This difference may be ascribed to an increase in the inhomogeneity
of spatial distribution of junctions in an elastomeric matrix caused
by the presence of filler particles.

\item
The characteristic time of relaxation, $\tau_{\ast}$, for the unfilled and
filled rubbers increases with the elongation ratio, $\lambda$.
This observation contradicts the strain--time superposition principle
conventionally employed for the analysis of glassy polymers.
An increase in $\tau_{\ast}$ with $\lambda$ may be explained
by the mechanically induced crystallization of natural rubber in the
regions with low concentration of junctions.
The rate of growth for the average relaxation time, $\tau_{\ast}$, with strains
is maximal for the unfilled rubber and it decreases with an increase
in the content of filler.

\item
The average energy for breakage of strands, $\Omega$, decreases with an increase
in the filler content, whereas the width of the Gaussian distribution of
energies, $\xi=\Sigma/\Omega$, is practically independent of the concentration
of carbon black.
The distribution function for the energies of breakage, $p(\omega)$, does not
depend on the elongation ratio, $\lambda$.

\item
The ratio of stresses in the regions with low concentration of junctions
to the total stress in a specimen, $A$, increases with $\lambda$.
This means that at uniaxial tension, stresses in the domains with low concentration
of cross-links grow more rapidly then in the elastomeric matrix and in the regions with
high concentration of junctions.
\end{enumerate}

\section*{Acknowledgements}

We would like to express our gratitude to Dr. K. Fuller (TARRC, UK)
for providing us with rubber specimens for the analysis
and to Prof. P. Haupt and Dr. S. Hartmann (University of Kassel, Germany)
for sending us an electronic version of their experimental data.
We are indebted to Mr. G. Seifritz for his help in carrying out
mechanical tests.
ADD acknowledges stimulating discussions with Prof. N. Aksel
(University of Bayreuth, Germany).

\newpage

\begin{center}
{\bf Table 1:} Chemical composition of rubbers (phr by weight) used
in the experimental study

\vspace*{6 mm}

\begin{tabular}{@{} l r r @{}}
\hline Formulation & EDS--16 & EDS--19 \cr \hline
Natural rubber         & 100 & 100\\
Zinc oxide             & 5   & 5  \\
Stearic acid           & 2   & 2  \\
Carbon black (N330)    & 45  & 1  \\
Process oil            & 4.5 & 0  \\
Antidegradant(HPPD)    & 3   & 3  \\
Antiozonant wax        & 2   & 2  \\
Accelerator (CBS)      & 0.6 & 0.6\\
Sulphur                & 2.5 & 2.5\\
\hline
\end{tabular}
\end{center}
\vspace*{30 mm}

\begin{center}
{\bf Table 2:} Adjustable parameters of the model
\vspace*{6 mm}

\begin{tabular}{@{} l c c c c c c c @{}}
\hline Rubber & $W$ & $\Sigma$ & $\xi$ & $A_{0}$ & $A_{1}$ & $\tau_{0}$ &
$\tau_{1}$ \cr \hline
Unfilled   & 11.70 & 6.10 & 0.52 & 0.20 & 0.054    & 0.00 &      9.20\\
R1         & 5.30 &  2.80 & 0.53 & 0.17 & 0.024    & 1.01 &      0.12\\
R2         & 5.13 &  3.18 & 0.62 & 0.18 & 0.026    & 0.68/1.30 & 3.02/0.30\\
R3         & 5.55 &  3.36 & 0.61 & 0.15/0.07 & $-0.07/0.02$ & 0.54/0.22 & 2.78/0.89\\
\hline
\end{tabular}
\end{center}

\newpage
\section*{List of figures}

{\bf Figure 1:} The dimensionless ratio $R$ versus time $t$ (s)
for unfilled rubber in relaxation tests with the 
elongation ratios $\lambda$ at room temperature. 
Symbols: experimental data. 
Solid lines: results of numerical simulation.
Curve 1: $\lambda=1.2$;
curve 2: $\lambda=1.4$;
curve 3: $\lambda=1.8$

\noindent
{\bf Figure 2:} The characteristic time of relaxation $\tau_{\ast}$ (s)
versus the first invariant of the Cauchy deformation tensor $I_{1}$ 
for unfilled rubber in relaxation tests at room temperature.
Circles: treatment of observations.
Solid line: approximation of experimental data by Eq. (59)
with $\tau_{0}=0$ and $\tau_{1}=9.2012$

\noindent
{\bf Figure 3:} The dimensionless parameter $A$ versus the first invariant 
of the Cauchy deformation tensor $I_{1}$ for unfilled rubber in 
relaxation tests at room temperature.
Circles: treatment of observations.
Solid line: approximation of experimental data by Eq. (59)
with $A_{0}=0.2028$ and $A_{1}=0.0537$

\noindent
{\bf Figure 4:} The dimensionless ratio $R$ versus time $t$ (s) for
CB filled rubber (R1) in relaxation tests with the elongation ratios
$\lambda$ at room temperature. Circles: experimental data.
Solid lines: results of numerical simulation.
Curve 1: $\lambda=1.2$;
curve 2: $\lambda=1.4$;
curve 3: $\lambda=1.8$;
curve 4: $\lambda=2.0$

\noindent
{\bf Figure 5:} The dimensionless ratio $R$ versus time $t$ (s) 
for CB filled rubber (R2) in relaxation tests 
with the elongation ratios $\lambda$ at room temperature. 
Circles: experimental data.
Solid lines: results of numerical simulation.
Curve 1: $\lambda=1.2$;
curve 2: $\lambda=1.4$;
curve 3: $\lambda=1.8$;
curve 4: $\lambda=2.0$

\noindent
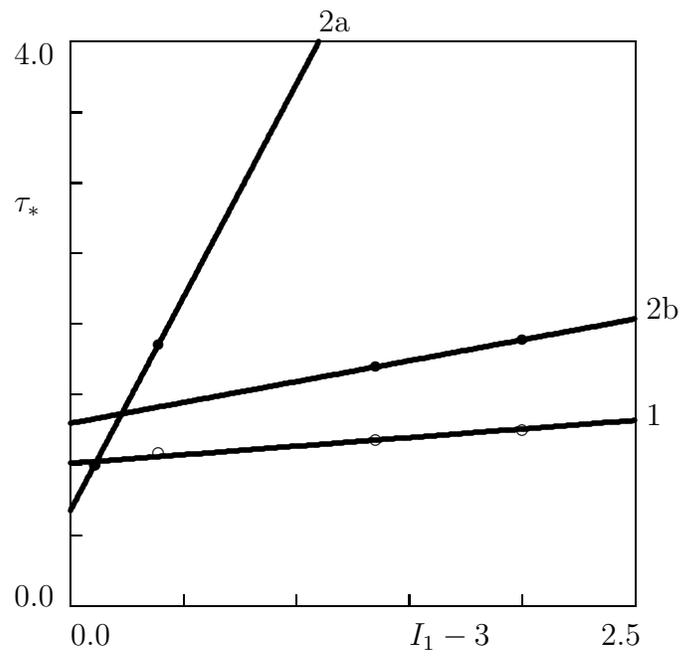
{\bf Figure 6:} The characteristic time of relaxation $\tau_{\ast}$ (s) versus 
the first invariant $I_{1}$ of the Cauchy deformation tensor for 
CB filled rubbers in relaxation tests at room temperature. 
Symbols: treatment of observations.
Unfilled circles: R1; filled circles: R2.
Solid lines: approximation of experimental data by Eq. (59).
Curve 1: $\tau_{0}=1.0114$, $\tau_{1}=0.1217$;
curve 2a: $\tau_{0}=0.6777$, $\tau_{1}=3.0218$;
curve 2b: $\tau_{0}=1.2954$, $\tau_{1}=0.2957$

\noindent
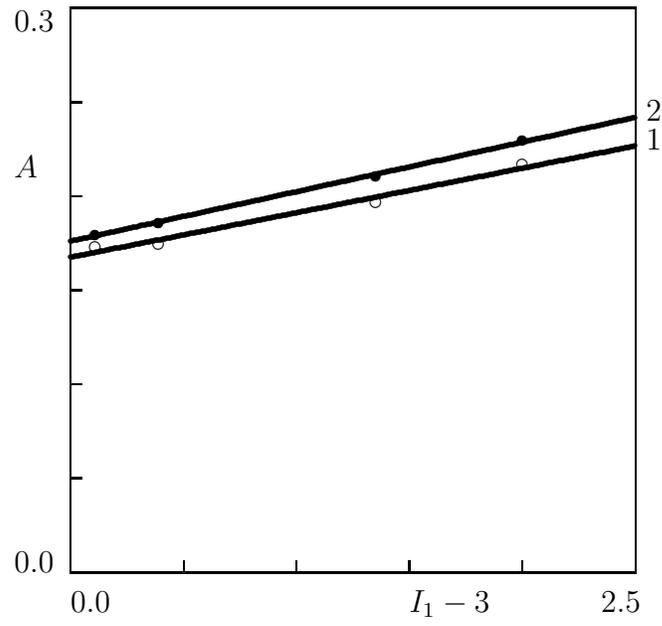
{\bf Figure 7:} The dimensionless parameter $A$ versus the first
invariant of the Cauchy deformation tensor $I_{1}$ for two
CB filled rubbers in relaxation tests at room temperature. 
Symbols: treatment of observations. 
Unfilled circles: R1;
filled circles: R2.
Solid lines: approximation of experimental data by Eq. (59).
Curve 1: $A_{0}=0.1677$, $A_{1}=0.0237$; 
curve 2: $A_{0}=0.1762$, $A_{1}=0.0263$

\noindent
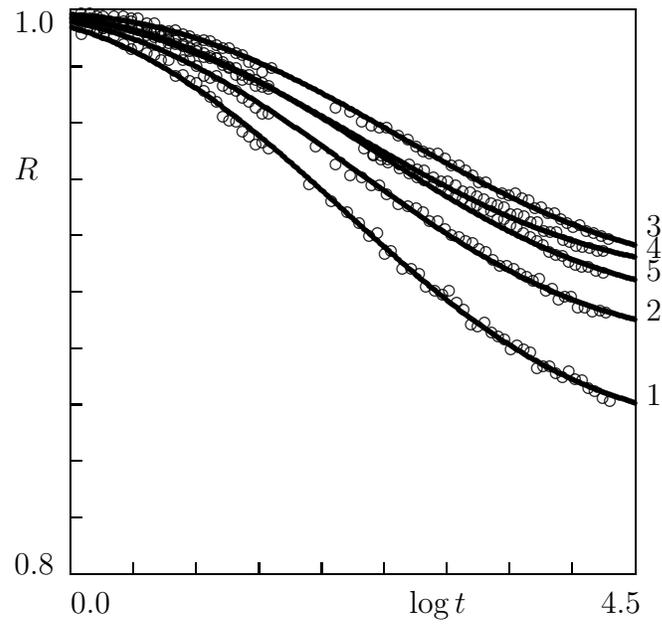
{\bf Figure 8:} The dimensionless ratio $R$ versus time $t$ (s) 
for CB filled rubber in relaxation tests with the elongation
ratios $\lambda$ at room temperature. 
Circles: experimental data \cite{HS01}.
Solid lines: results of numerical simulation.
Curve 1: $\lambda=1.19$; 
curve 2: $\lambda=1.38$;
curve 3: $\lambda=1.58$;
curve 4: $\lambda=1.76$;
curve 5: $\lambda=1.92$

\noindent
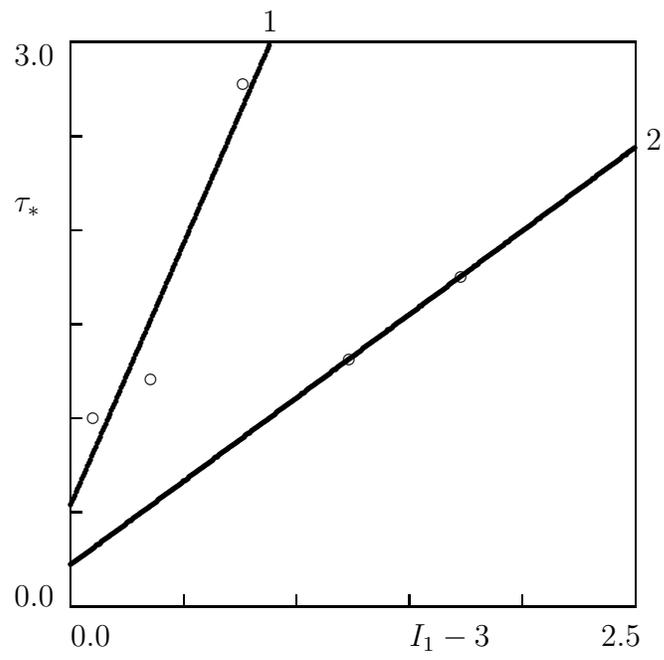
{\bf Figure 9:} The characteristic time of relaxation $\tau_{\ast}$ (s)
versus the first invariant of the Cauchy deformation tensor $I_{1}$ 
in relaxation tests at room temperature.
Circles: treatment of observations \cite{HS01}.
Solid lines: approximation of the experimental data by Eq. (59).
Curve 1: $\tau_{0}=0.5373$, $\tau_{1}=2.7795$;
curve 2: $\tau_{0}=0.2204$, $\tau_{1}=0.8877$

\noindent
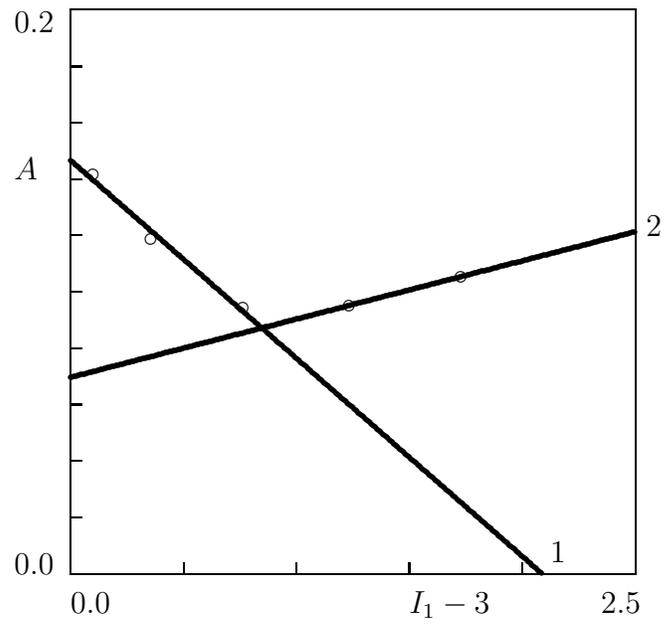
{\bf Figure 10:} The dimensionless parameter $A$ versus the first invariant
of the Cauchy deformation tensor $I_{1}$ for relaxation tests 
at room temperature.
Circles: treatment of observations \cite{HS01}.
Solid lines: approximation of the experimental data by Eq. (59).
Curve 1: $A_{0}=0.1467$, $A_{1}=-0.0702$;
curve 2: $A_{0}=0.0697$, $A_{1}=0.0206$

\setlength{\unitlength}{0.75 mm}
\begin{figure}[h]
\begin{center}

\end{center}
\vspace*{4 mm}

\caption{}
\end{figure}

\begin{figure}[tbh]
\begin{center}
%
\end{center}
\vspace*{4 mm}

\caption{}
\end{figure}

\begin{figure}[h]
\begin{center}
%
\end{center}
\vspace*{4 mm}

\caption{}
\end{figure}

\begin{figure}[h]
\begin{center}
%
\end{center}
\vspace*{4 mm}

\caption{}
\end{figure}

\begin{figure}[h]
\begin{center}
%
\end{center}
\vspace*{4 mm}

\caption{}
\end{figure}
\end{document}